\documentclass[aip,apl,preprint]{revtex4}
\usepackage{graphicx}
\usepackage{amsmath}
\usepackage{stmaryrd}
\usepackage{cellspace}

\begin{document}
\title{Gate controlled coupling of intersubband plasmons}
\author{A. Delteil}
\affiliation{Universit\'e Paris Diderot, Sorbonne Paris Cit\'e, Laboratoire Mat\'eriaux et Ph\'enom\`enes Quantiques, UMR7162, 75013 Paris, France}
\author{A. Vasanelli}
\email{angela.vasanelli@univ-paris-diderot.fr}
\affiliation{Universit\'e Paris Diderot, Sorbonne Paris Cit\'e, Laboratoire Mat\'eriaux et Ph\'enom\`enes Quantiques, UMR7162, 75013 Paris, France}
\author{Y. Todorov}
\affiliation{Universit\'e Paris Diderot, Sorbonne Paris Cit\'e, Laboratoire Mat\'eriaux et Ph\'enom\`enes Quantiques, UMR7162, 75013 Paris, France}
\author{B. Paulillo}
\affiliation{Universit\'e Paris Diderot, Sorbonne Paris Cit\'e, Laboratoire Mat\'eriaux et Ph\'enom\`enes Quantiques, UMR7162, 75013 Paris, France}
\author{G. Biasiol}
\affiliation{IOM CNR, Laboratorio TASC, Area Science Park, I-34149 Trieste, Italy}
\author{L. Sorba}
\affiliation{NEST, Istituto Nanoscienze-CNR and Scuola Normale Superiore, I-56127 Pisa, Italy}
\author{C. Sirtori}
\affiliation{Universit\'e Paris Diderot, Sorbonne Paris Cit\'e, Laboratoire Mat\'eriaux et Ph\'enom\`enes Quantiques, UMR7162, 75013 Paris, France}

\date{\today}

\begin{abstract}
The optical response of a heavily doped quantum well, with two occupied subbands, has been investigated as a function of the electronic density. It is shown that the two optically active transitions are mutually coupled by dipole-dipole Coulomb interaction, which strongly renormalizes their absorption amplitude. In order to demonstrate this effect, we have measured a set of optical spectra on a device in which the electronic density can be tuned by the application of a gate voltage. Our results show that the absorption spectra can be correctly described only by taking into account the Coulomb coupling between the two transitions. As a consequence, the optical dipoles originating from intersubband transitions are not independent, but rather coupled oscillators with an adjustable strength. 
\end{abstract}

\maketitle
Intersubband (ISB) optical excitations in a semiconductor two-dimensional  electron gas are a collective phenomenon, that involves modes of charge density oscillations, the ISB plasmons~\cite{chen, allen, ando, wendler}. When electrons populate only the first subband, the signature of the collective modes appears in the optical response as a blue shift of the absorption resonance with respect to the energy separation between the bare states, the well-known depolarization shift~\cite{tsujino, luin}. In a symmetric potential, if the electronic density in the conduction band is such that two subbands are occupied, the only dipole allowed transitions are the $1 \rightarrow 2$ and the $2 \rightarrow 3$. In this case the collective character of the ISB absorption manifests itself in a redistribution of the oscillator strength from the low energy to the high energy mode~\cite{warburton_PRL, zeller}. In this work we demonstrate that this redistribution is a direct consequence of the dipole-dipole Coulomb coupling between the ISB plasmons associated to the two transitions. To this end, absorption spectra of a quantum well with two occupied subbands are measured in a device in which the electronic density can be adjusted by a gate voltage. We show that the coupling between the plasmons depends on the electronic density in the well and results in a hybridization of the two ISB transitions into new normal modes of the system. Our data can be illustrated using a quantum model~\cite{todorov_PRB2012} that very well reproduces the energy peaks and their coupling with the infrared radiation.

\begin{figure}[ht]
\centering
\includegraphics[width=0.8\columnwidth]{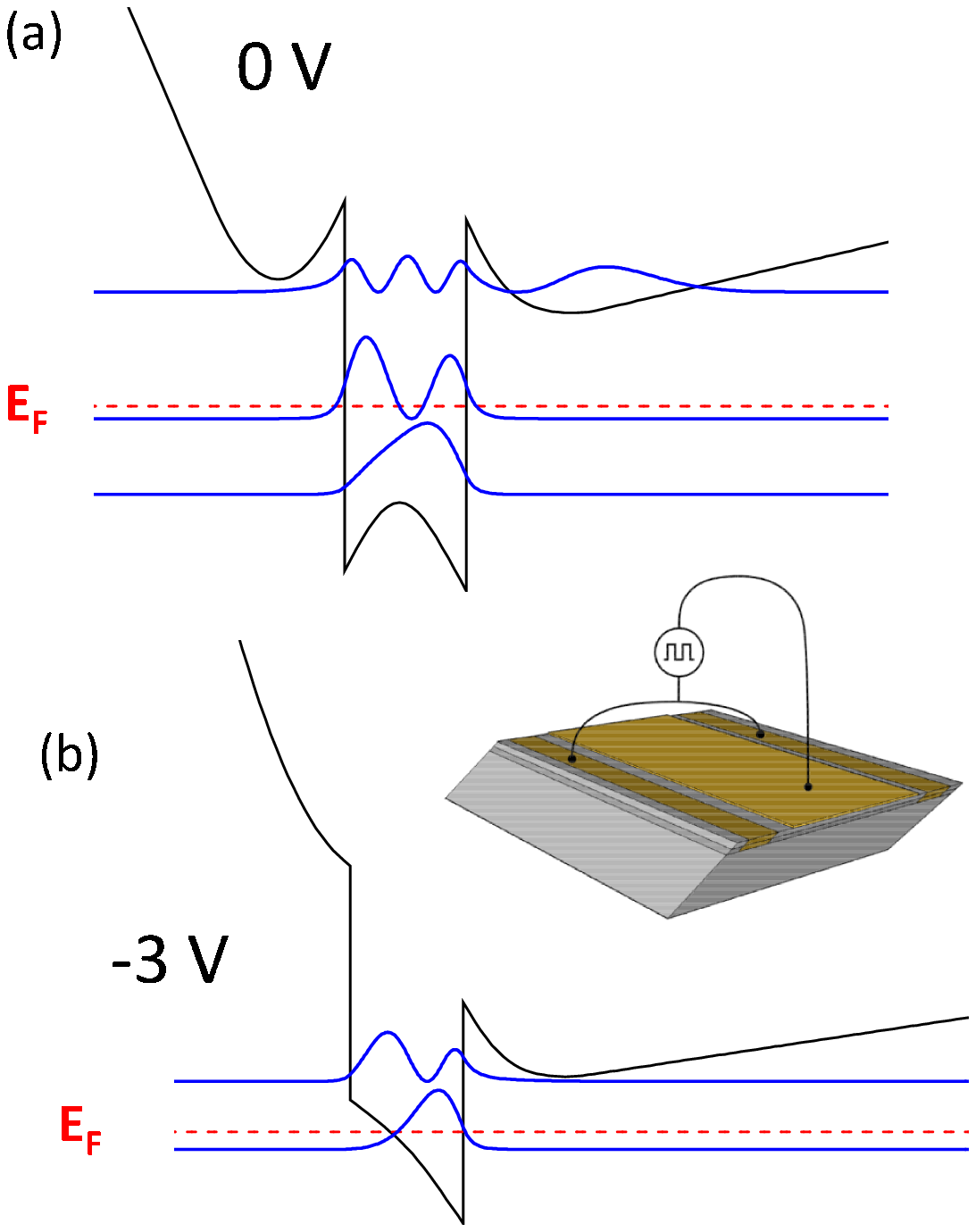}
\caption{Band diagram of the structure at 77K, as calculated with a self-consistent Schr\"{o}dinger - Poisson solver at 0V (panel a) and at -3V (panel b). The square moduli of the relevant wave functions are plotted in (blue) continuous lines and the Fermi energy is indicated by a (red) dashed line. The inset sketches the geometry of the device used to perform absorption measurements under the application of a gate voltage.}
\label{Fig1}
\end{figure}

\begin{figure}[ht]
\centering
\includegraphics[width=0.7\columnwidth]{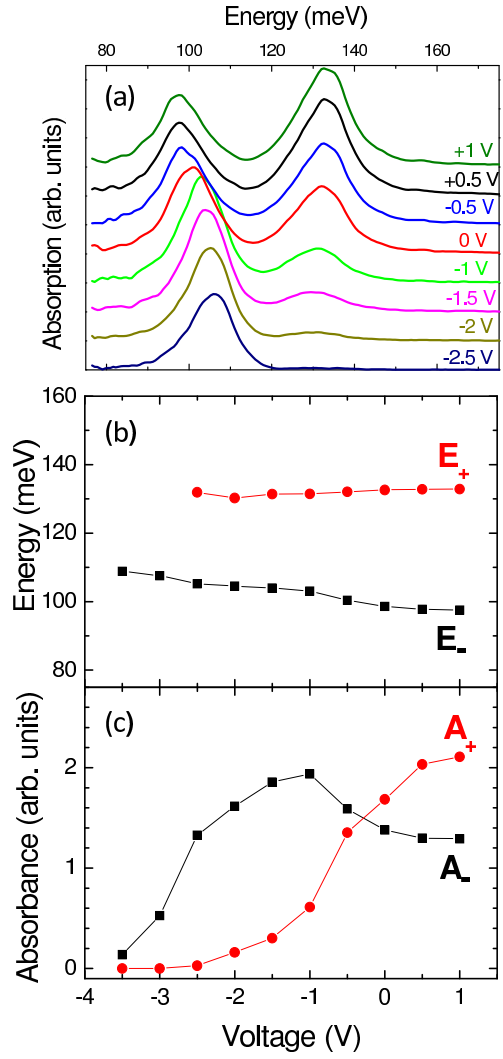}
\caption{a) Absorption spectra measured at 77K for different gate voltages. b) Energy position of the absorption peaks as a function of the applied voltage. c) Area associated to the two absorption peaks as a function of the applied voltage.}
\label{Fig2}
\end{figure}

The sample used for this study is a single GaAs/Al$_{0.45}$Ga$_{0.55}$As quantum well, 11 nm thick, grown by molecular beam epitaxy. The quantum well is Si doped with a donor concentration of $6.6 \times 10^{18}$ cm$^{-3}$. The Al$_{0.45}$Ga$_{0.55}$As barriers are also Si doped ($3.5 \times 10^{18}$ cm$^{-3}$) within a thickness of 10 nm, and the Si layers are offset by about 1 nm from the well. The band diagram of the structure at 77K, as calculated with a self-consistent Schrödinger - Poisson solver~\cite{snider}, is shown in figure~\ref{Fig1}a, together with the square moduli of the wave functions. The electron density in the quantum well is $3.3 \times 10^{12}$ cm$^{-2}$ and the Fermi level, indicated in the figure by a (red) dashed line, is above the second subband. In order to vary continuously the electron density in the sample, a Ti/Au gate was evaporated on the surface and ohmic NiAuGe/NiAu contacts were diffused down to the quantum well layer. The effect of the application of a negative gate voltage is to progressively deplete the quantum well. As an example, figure \ref{Fig1}b presents the simulated band diagram of the structure when a voltage of -3V is applied to the device. In this case only the first subband is occupied. The quantum well is completely depleted for -5V bias. For optical measurements, we fabricated $45^\circ$-edge multipass waveguides, as sketched in the inset of figure~\ref{Fig1}. Absorption spectra were acquired with an infrared Fourier transform spectrometer in step-scan mode and using infrared light polarized perpendicularly to the QW layers and propagating under the gate. The gate voltage was square-wave modulated between -8 V and the desired voltage, at a frequency of 10kHz~\cite{luin}. The obtained spectra are subsequently normalized by a background collected at $V_g=-8$~V. The spectra measured between -2.5 V and +1 V at 77K are shown in figure~\ref{Fig2}a. They present a single peak when only the fundamental subband is occupied. As soon as the Fermi level is above the second subband, two resonances are visible in the spectra, whose energy position and amplitude depend on the applied voltage. Figure~\ref{Fig2}b presents the energy position of the observed absorption maxima as a function of the applied bias. When going from -2V to +1V, the energy separation between the high energy peak ($E_+$) and the low energy one ($E_-$) increases of 10 meV. In figure ~\ref{Fig2}c we also plot as a function of the voltage the area associated to each absorption peak, which is proportional to the absorption amplitude. The amplitude of the low energy peak, $A_-$, increases monotonously with the voltage, up to -1V, where it starts to decrease. At the same time the amplitude of the high energy peak, $A_+$, rapidly increases and becomes comparable to $A_-$ at $V_g=0$~V. Note that this is in contrast to what expected by considering the ISB absorption as a single particle process that would give a ratio between the areas under the two peaks close to 3. Furthermore, the amplitude of the $1 \rightarrow 2$ transition should saturate instead of decreasing when the Fermi level is above the second subband, as the population difference ($N_1-N_2$) stays constant~\cite{nota1}. This experimental behavior demonstrates that the two resonances observed in the absorption spectra are not simply the $1 \rightarrow 2$ and $2 \rightarrow 3$ transitions, but new eigenmodes of the system. In the following we will demonstrate that this renormalization is issued from the dipole-dipole Coulomb coupling between the ISB plasmons associated to the two transitions.

\begin{figure}[ht]
\centering
\includegraphics[width=0.7\columnwidth]{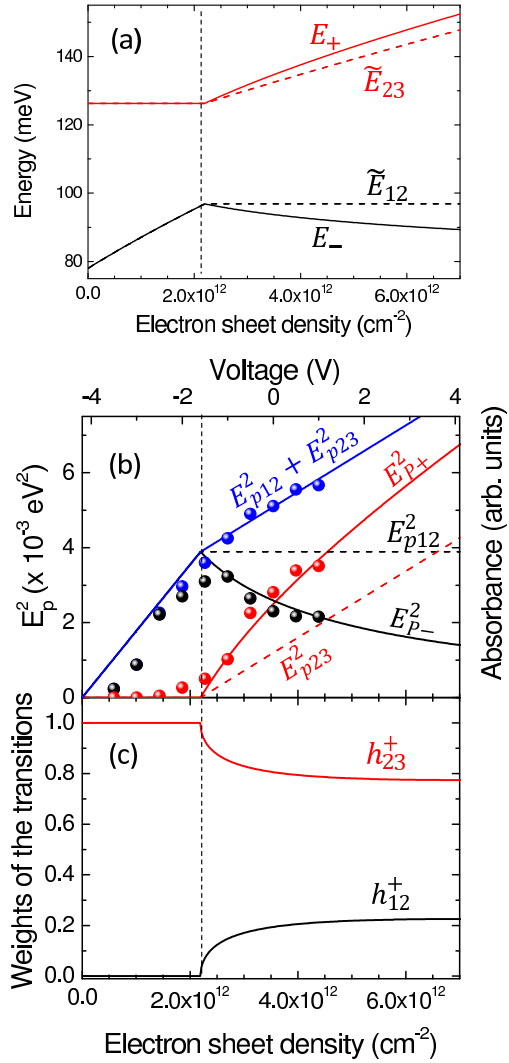}
\caption{a) Normal modes of the system of coupled plasmons (continuous lines) as a function of the electronic density in the well. They are compared to the uncoupled plasmon energies (dashed lines). b) Squared plasma energies calculated by using coupled (continuous lines) or uncoupled (dashed lines) plasmons as a function of the electronic density. The experimental absorbance as a function of the voltage is reported as bullets. c) Weights of the $1 \rightarrow 2$ and $2 \rightarrow 3$ transitions on the high energy mode $E_+$.}
\label{Fig3}
\end{figure}

The optical response of the two dimensional electron gas is studied in the framework of the dipole gauge Hamiltonian~\cite{todorov_PRB2012}, which contains the effects of the dipole-dipole interactions and the coupling of the electronic polarization with light. For the high electronic densities considered here the exchange-correlation  part of the Coulomb interaction can be neglected~\cite{ando}. The matter part of the dipole gauge Hamiltonian, $H_{\rm{matter}}$, can be written in terms of bosonic operators $p_{j,j+1}^\dagger$ (respectively $p_{j,j+1}$) corresponding to the creation (resp. annihilation) of plasmonic excitations between the subbands $j$ and $j+1$:
\begin{equation}
H_{\rm{matter}}=\widetilde{E}_{12} p_{12}^\dagger p_{12} + \widetilde{E}_{23} p_{23}^\dagger p_{23} + \hbar \Xi_{12,23} \left( p_{12}^\dagger + p_{12} \right) \left( p_{23}^\dagger + p_{23} \right)
\label{Hmatter}
\end{equation}
The energies associated to the plasmonic excitations are: $\widetilde{E}_{j,j+1}=\sqrt{E_{j,j+1}^2+ E_{P_{j,j+1}}^2}$, where $E_{j,j+1}$ is the energy separation between subbands $j+1$ and $j$, while $E_{P_{j,j+1}}=\hbar e \sqrt{\frac{f_{j,j+1} \left( N_j -N_{j+1} \right)}{m^* \epsilon_0 \epsilon_{st} L_{\rm{eff}}}}$ is the plasma energy of the transition. Here $N_j$ is the electronic density into the subband $j$, $f_{j,j+1}$ is the oscillator strength of the transition $j \rightarrow j+1$, $\epsilon_{st}$ is the static background dielectric constant, $m^*$ the constant effective mass and $L_{eff}$ is the effective quantum well thickness renormalized by the Coulomb interaction~\cite{ando}. The last term in eq.~\ref{Hmatter} describes the interaction between the ISB plasmons, with a coupling constant $\hbar \Xi_{12,23}=\frac{1}{2} C_{12,23} \frac{E_{P_{12}}E_{P_{23}}}{\sqrt{{\widetilde{E}_{12}}\widetilde{E}_{23}}}$. The quantity $C_{12,23}$ is a coefficient describing the overlap between the microscopic currents associated to the two ISB transitions~\cite{todorov_PRB2012}. This coefficient is very close to 1 in a square quantum well, and equal to 0.92 at 0V, in our system. This gives a value of the coupling constant $\hbar \Xi_{12,23} = 10$ meV. It is important to underline that $\hbar \Xi_{12,23}$  directly depends on the electronic densities in the subbands, via the plasma frequencies and therefore it can be tuned in our experimental device by changing the gate voltage. 
The coupling between the ISB plasmons gives rise to new eigenmodes of the system that can be found by diagonalising $H_{\rm{matter}}$. For this, we introduce new bosonic operators $P_+$ and $P_-$, linear combinations of $p_{12}$, $p_{23}$ and their hermitian conjugates:
\begin{equation}
\label{transform}
\left( \begin{array}{c} P_+ \\ P_- \end{array} \right) = Y \left( \begin{array}{c} p_{12} \\ p_{23} \end{array} \right) + T \left( \begin{array}{c} p_{12}^\dagger \\ p_{23}^\dagger \end{array} \right)
\end{equation}
With this transformation, it is possible to show that the normal modes of the system, $E_\pm$, associated to the bosonic operators $P_+$ and $P_-$, result from the solution of the following secular equation:
\begin{equation}
\label{autovalori}
\left( \widetilde{E}_{12}^2 - E^2 \right) \left( \widetilde{E}_{23}^2 - E^2 \right)-4\widetilde{E}_{12} \widetilde{E}_{23} {\left( \hbar \Xi_{12,23} \right)}^2=0
\end{equation}
Note that this equation is formally identical to that describing the ultra-strong coupling between a cavity mode and an ISB excitation~\cite{todorov_PRL2010}, with the Rabi energy replaced by the coupling constant between the two ISB plasmons. Figure~\ref{Fig3}a shows the eigenenergies $E_\pm$ (continuous lines) as a function of the electronic sheet density in the well. The calculation has been performed by using the energies and wavefunctions calculated at 0V (presented in fig.~\ref{Fig1}a), without considering their variation introduced by the applied voltage. The eigenenergies of the coupled system are compared to the energies $\widetilde{E}_{12}$ and $\widetilde{E}_{23}$ of the uncoupled ISB plasmons (dashed lines) that show only the usual depolarization shift. We can see that when the excited subband is occupied, the coupling increases the separation of the two modes $E_\pm$ compared to the uncoupled plasmons. Nevertheless, this difference cannot be used to discriminate the existence of the coupling between the plasmons, as it can be strongly influenced by the band-bending due to the applied voltage.

The effect of the coupling is much more visible in the dependence of the absorbance (the area under the absorption peak) as a function of the electronic density. This is illustrated in fig.~\ref{Fig2}c, where one can observe that, above a certain density, the area under the low energy peak reduces for increasing electronic concentration. This effect is associated with a redistribution of the oscillator strength between the two normal modes that arise from the increasing coupling of the transitions. A redistribution of the absorbance was already observed in ref.~\onlinecite{warburton_PRL} by varying the electronic density on the subbands as a function of the temperature. In our work we have varied through a gate the carrier density, thus the coupling strength between the transitions. The results of our measurement show that the oscillator strength redistribution is associated with the formation of the normal modes. Their interaction with light is mediated by effective plasma energies obtained as~\cite{todorov_PRB2012_2, delteil}:
\begin{equation}
E_{P_{+(-)}}=E_{P_{12}} X_{12}^{+(-)} \sqrt{\frac{E_{+(-)}}{\widetilde{E}_{12}}} + E_{P_{23}} X_{23}^{+(-)} \sqrt{\frac{E_{+(-)}}{\widetilde{E}_{23}}}
\end{equation}
with $X_{j,j+1}^{+(-)}$ the matrix elements of $X=(Y+T)^{-1}$. The absorption coefficient can therefore be expressed as a function of the squared plasma energies, like in the semi-classical Drude - Lorentz oscillator model. Figure~\ref{Fig3}b presents (continuous lines) the square of the normal mode plasma energies $E_{P_{+(-)}}^2$ (respectively in red and black) as a function of the electronic density. They are compared to the squared plasma energies $E_{P_{12}}^2$ and $E_{P_{23}}^2$ (dashed lines) that one would have obtained without considering the coupling between ISB plasmons. The effect of the coupling is a redistribution of the absorption amplitude from the low energy to the high energy mode, without changing the total absorption. We verify indeed that $E_{P_{12}}^2 + E_{P_{23}}^2 = E_{P_{+}}^2 + E_{P_{-}}^2$. This redistribution is responsible for the reduction of the amplitude associated to the low energy peak. The experimental absorption amplitudes as a function of the voltage (bullets), already shown in fig.~\ref{Fig2}c, are now reported on the simulation graph, Fig.~\ref{Fig3}b. To calibrate the two horizontal scales, we have used the value of the electronic density at 0V and the voltage at which the Fermi level goes below the second subband ($\approx -2.5$V). The graph also reports the total absorption (in blue) obtained from both simulations and experiments. There is very good agreement between the experimental and simulated behavior as a function the electronic density (gate voltage), demonstrating that the optical response of our device is controlled by the coupling between the ISB plasmon. 

As the new eigenmodes of the system are linear combinations of the two ISB plasmons, we can define two coefficients $h_{12}^{+(-)}$ and $h_{23}^{+(-)}$, giving the weight of each ISB plasmon (respectively 1-2 and 2-3) to the eigenenergies ($E_+$ and $E_-$) of the coupled system. They are the analogue for the Coulomb coupling between plasmons of the Hopfield coefficients for the light-matter interaction~\cite{Hopfield}. The contributions of the two ISB excitations to the mode $E_+$ can be defined as: $h_{12}^+ = {\left \vert Y_{12}^+ \right \vert}^2 - {\left \vert T_{12}^+ \right \vert}^2$ and $h_{23}^+ = {\left \vert Y_{23}^+ \right \vert}^2 - {\left \vert T_{23}^+ \right \vert}^2$ where $Y_{j,j+1}^+$ and $T_{j,j+1}^+$ are the matrix elements appearing in eq.~\ref{transform}. The weights of two transitions into the high energy normal mode, $E_+$, are plotted in fig.~\ref{Fig3}c as a function of the electronic density. Note that the hybridization between the two plasmons becomes effective soon after the Fermi level is above the second subband (for an electronic density of $2.1 \times 10^{12}$ cm$^{-2}$). The contribution of the 2-3 transition in the high energy mode rapidly saturates at a value of $77\%$ for our specific structure. In general this value depends on the detuning between the two transition energies. The hybridization is maximal ($h_{12}^{+(-)} = h_{23}^{+(-)} = 1/2$) for a parabolic quantum well, where the detuning between the transition energies vanishes. The result shown in fig.~\ref{Fig3}c underlines that the high energy peak contains a non-negligible contribution from the $1 \rightarrow 2$ ISB plasmon, meaning that the high energy mode cannot be attributed to the $2 \rightarrow 3$ transition only. In a square quantum well with several occupied subbands, an equal number of normal modes are issued from the Coulomb coupling between ISB transitions, and the high energy one concentrates the entire absorption amplitude of the system, as discussed in ref.~\onlinecite{delteil}.

\begin{figure}[ht]
\centering
\includegraphics[width=0.7\columnwidth]{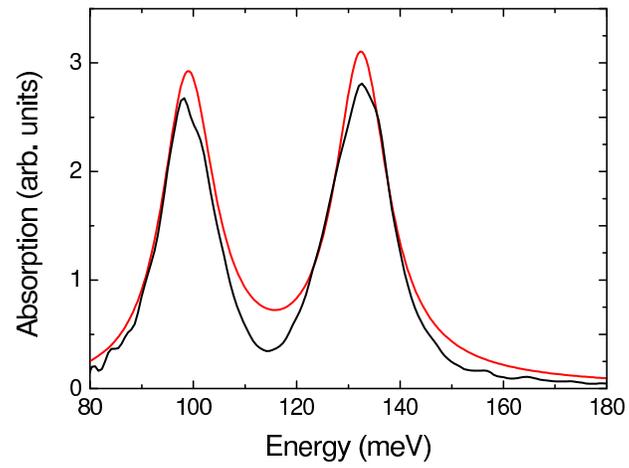}
\caption{Simulated absorption spectrum at 0V (red line), compared to the experimental one (black line).}
\label{Fig4}
\end{figure}

The optical spectra can be simulated by deriving a dielectric constant associated to our system, $\epsilon_{zz}(E)$, as:
\begin{equation}
\label{epsilon}
\frac{\epsilon_{st}}{\epsilon_{zz}\left( E \right)}=1+ \frac{E_{P_+}^2}{E^2-E_+^2+i E \gamma} + \frac{E_{P_-}^2}{E^2-E_-^2+i E \gamma}
\end{equation}
This expression is identical to that obtained in a semiclassical model~\cite{warburton_PRB} starting from a dielectric function of the form: $\frac{\epsilon_{zz}\left( E \right)}{\epsilon_{st}}=1- \frac{{E_{P_{12}}}^2}{E^2-{E_{12}}^2+i E \gamma} + \frac{{E_{P_{23}}}^2}{E^2-{E_{23}}^2+i E \gamma}$.  The absorption per unit area can in this case be calculated as the imaginary part of $\frac{\epsilon_{st}}{\epsilon_{zz}}$.~\cite{ando} Figure \ref{Fig4} presents the simulated absorption spectrum at 0V, compared to the experimental one. The agreement between the two spectra is excellent, showing that the redistribution of the amplitude between the peaks is correctly reproduced by our model. 

In conclusion, we demonstrated that the optical response of a highly doped two-dimensional electron gas depends on the coupling between ISB plasmons, which is a function of the electronic density in the subbands, controlled by a gate voltage. The experimental results are very well explained by considering the ISB plasmons as quantum harmonic oscillators, mutually coupled by the dipole-dipole part of the Coulomb interaction. Within this framework, the absorption energies are identified as the new normal modes of the system, and the absorption amplitudes are directly related to effective plasma frequencies. Our findings indicate that the coherences induced by Coulomb interaction in dense electron gases offer new degrees of freedom for the realization of active media with electrically tunable optical properties~\cite{miao}.

\begin{acknowledgments}
This work has been partially supported by the French National Research Agency (ANR) in the frame of its Nanotechnology and Nanosystems program P2N, Project No. ANR-09-NANO-007. We acknowledge financial support from the ERC grant ``ADEQUATE''.
\end{acknowledgments}

\newpage

\end{document}